# A Fourfold Pathogen Reference Ontology Suite

John Beverley[1,2*], Shane Babcock[1,3], Carter Benson[2,4], Giacomo De Colle[1,2], Sydney Cohen[1], Alexander D. Diehl[1,2], Ram A.N.R. Challa[2], Rachel A. Mavrovich[1,2], Joshua Billig[1,2], Anthony Huffman[5], and Yongqun He[5]

[1]*National Center for Ontological Research, United States*
[2]*University at Buffalo, State University of New York, United States*
[3]*KadSci LLC, United States*
[4]*CUBRC, Inc., United States*
[5]*University of Michigan-Ann Arbor, United States*

**\*Corresponding Author**
johnbeve@buffalo.edu

## Abstract

**Background**: Infectious diseases remain a critical global health challenge, and the integration of standardized ontologies plays a vital role in managing related data. The Infectious Disease Ontology (IDO) and its extensions, such as the Coronavirus Infectious Disease Ontology (CIDO), are essential for organizing and disseminating information related to infectious diseases. The COVID-19 pandemic highlighted the need for updating IDO and its virus-specific extensions. There is an additional need to update IDO extensions specific to bacteria, fungus, and parasite infectious diseases.
**Methods**: The "hub-and-spoke" methodology is adopted to generate pathogen-specific extensions of IDO: Virus Infectious Disease Ontology (VIDO), Bacteria Infectious Disease Ontology (BIDO), Mycosis Infectious Disease Ontology (MIDO), and Parasite Infectious Disease Ontology (PIDO).
**Results**: IDO is introduced before reporting on the scopes, major classes and relations, applications and extensions of IDO to VIDO, BIDO, MIDO, and PIDO.
**Conclusions**: The creation of pathogen-specific reference ontologies advances modularization and reusability of infectious disease ontologies within the IDO ecosystem. Future work will focus on further refining these ontologies, creating new extensions, and developing application ontologies based on them, in line with ongoing efforts to standardize biological and biomedical terminologies for improved data sharing, quality, and analysis.

## Background

Infectious diseases are a persistent and evolving challenge to global health. Research communities produce enormous amounts of data regarding various aspects of infectious diseases. Ontologies – controlled vocabularies of terms and relationships among them - play a critical role in cleaning, disseminating, and coordinating such information by encoding them into standardized machine-interpretable languages [1]. The Open Biological and Biomedical Ontology (OBO) Foundry has long-provided useful guidelines around ontology development among life science researchers [2]. Accordingly, numerous infectious disease ontologies have been created and added to the OBO Foundry library (http://obofoundry.org/). Of note is the Infectious Disease Ontology (IDO; https://bioportal.bioontology.org/ontologies/IDO) [3, 4], which serves as a starting point for more specific infectious disease ontologies in the OBO Foundry, such as the Coronavirus Infectious Disease Ontology (CIDO; https://bioportal.bioontology.org/ontologies/CIDO) [5, 6] and Influenza Infectious Disease Ontology (IDOFLU; https://bioportal.bioontology.org/projects/IQ) [7].

The height of the coronavirus pandemic witnessed a need to revisit the curation of IDO and its various extensions. This owed in part to revisions made to the terminological foundation upon which IDO terms and relations are based: the Basic Formal Ontology (BFO; https://bioportal.bioontology.org/ontologies/BFO). BFO is a top-level ontology covering general classes such as **material entity**,[1] **quality**, **process**, and **role** [1, 8, 9, 10], which provides a common starting point for many OBO Foundry ontologies. Shortly before the pandemic, BFO became the first top-level ISO/IEC 21838-1 standard (https://www.iso.org/standard/74572.html), with the standardization process resulting in adjustments to its core vocabulary. A handful of terms and relations in IDO were impacted by these upstream adjustments, which spurred action to update IDO, as well as its infectious disease extensions, with a particular emphasis on those extensions designed to represent viruses. Subsequent updates to IDO [3] and an evaluation of the work needed to align its extensions [11], led to the recognition that many such extension ontologies introduced terms and relations broader than the stated scope of their home ontology. For example, IDOFLU introduced the term **antiviral drug** – "A drug used specifically for treating viral infections" – which is not limited to the domain of influenza, but would indeed be useful for, say, the HIV Ontology

---

[1] We adopt the convention of placing in bold reference to classes and relations.

*A Fourfold Pathogen Reference Ontology Suite*(IDOHIV) [12], the Dengue Fever Ontology (IDODEN) [13], the Combined Ontology for Inflammatory Diseases (COID) [14], or the COVID-19 Ontology (CODO) [15].

In the interest of partially addressing such issues, the Virus Infectious Disease Ontology (VIDO; https://bioportal.bioontology.org/ontologies/VIDO) [16] was developed as an reference ontology [1] extension of IDO to the domain of virus-specific infectious diseases; it was intended to provide "guardrails" to researchers so that they may reap the benefits of aligning with more general ontologies like IDO without having to spend time modeling from the "top-down" [17]. In other words, VIDO was designed to provide researchers with vetted ontological representations closer to relevant virus infectious disease data. Subsequent alignment work emerging from the development of VIDO can be seen, for example, in the CIDO [5, 18].

Viruses are not, of course, the only infectious pathogens worthy of ontological representation; many extensions of IDO concern other pathogen types, such as bacteria and parasites. Following the "hub-and-spoke" advocated by initiators of IDO [19], teams of ontology engineers and subject-matter experts have pursued the creation of reference ontology extensions of IDO covering other pathogen types. Inspired by traditional classifications of pathogens, the resulting suite of reference ontologies includes VIDO as discussed, but also three new pathogen infectious disease reference ontologies: the Bacteria Infectious Disease Ontology (BIDO; https://github.com/infectious-disease-ontology-extensions/BIDO), the Mycosis Infectious Disease Ontology (MIDO; https://github.com/infectious-disease-ontology-extensions/MIDO), and the Parasite Infectious Disease Ontology (PIDO; https://github.com/infectious-disease-ontology-extensions/PIDO). As with VIDO, these reference ontologies will serve as intermediate layers between IDO and its extensions into more specific pathogen representations. Developers of these projects maintain that this layered approach – curating classes and relations common to a range of IDO extensions falling within these pathogen

types - will encourage modularity, improve ontology quality, and promote reuse as vetted ontological representations will more clearly connect to actual infectious disease data. In this respect, we take this project to be in the spirit of and compatible with the Core Ontology for Biology and Biomedicine (COB) effort, which has the admirable aim of coordinating well-developed terminologies across the OBO Foundry, for ease of access, reuse, and evaluation [20].

The following section provides an overview of the hub-and-spoke model leveraged by each extension to ensure alignment with IDO, as well as an overview of IDO terms and relations relevant to pathogen-type ontological representations. Subsequent sections introduce each reference ontology of the suite, with discussions of major classes and relations, design patterns, applications and extensions. We close by discussing limitations and future work.

**Methods**

**OWL, Protégé, and Reasoners.** Each ontology is represented in the OWL 2 Web Ontology Language (https://www.w3.org/TR/owl2-overview/). OWL is a decidable fragment of first-order logic, which is to say there exists an algorithm that can determine the truth-value for any statement expressed in the language in a finite number of steps. Restricting expressions to a decidable language allows automated consistency and satisfiability checking. Ontologies in the suite were developed using the Protégé editor (https://protege.stanford.edu/) and tested against automated reasoners such as HermiT [21] and Pellet [22].

**Ontology Engineering Best Practices.** Each ontology in the suite was designed to align with OBO Foundry best practices, thereby supporting interoperability with existing Foundry ontologies [2]. OBO Foundry design principles require that ontologies use a well-specified syntax, leverage an unambiguous common space of identifiers, be openly available in the public domain, be modularly developed with a specified scope, and import a common set of relations from the Relations Ontology (RO; https://obofoundry.org/ontology/ro.html). Development of each ontology, moreover, follows metadata conventions adopted by many OBO Foundry ontologies [23], such as requiring that any term introduced into the ontology has a unique identifier, textual definition, definition source, designation of term editor(s), and preferred term label.

Ontology developers in this project adopted the "hub and spokes" methodology [1, 3, 16] for ontology development. For example, VIDO would be considered a "spoke" extending from the IDO "hub". Extensions of IDO covering specific infectious diseases are created, first, by importing needed terms from IDO and other OBO Foundry ontologies, and second, by constructing the domain-specific terms where needed to adequately characterize the relevant domain. Intimately connected with the "hub and spokes" methodology is the adherence to extending definitions from existing ontologies following the "A is a B that C's" [1, 24] pattern. For example, the IDO class **appearance of disorder** (A) is defined as a '**process** (B) by which a **disorder** comes into existence (C)'[2] where the BFO class **process** is the parent class for **appearance of disorder**, while the remainder of the definition (C) differentiates **appearance of disorder** from sibling classes falling under **process**, such as **process of establishing an infection**, or **infectious disease epidemic**.

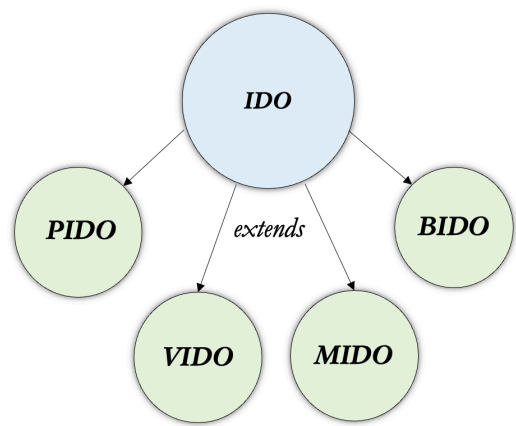

*Figure 1*: *The structure of the hub-and-spoke model used to develop downwards from IDO*

In the interest of coordinating development with existing OBO ontologies, developers imported terms where possible from existing OBO Foundry ontologies, such as the Gene Ontology

---
[2] "(A)", "(B)" and "(C)" are added here for illustrative purposes.

*A Fourfold Pathogen Reference Ontology Suite*

(GO) [25], the Protein Ontology (PRO) [26], the Ontology for Biomedical Investigations (OBI) [27], and of course IDO. **Table 1** highlights important terms leveraged from existing OBO Foundry ontologies, emphasizing content reused from IDO as it is the hub from which reference ontology spokes in the suite extend. Limited overviews for these terms will emerge from the discussion to follow as needed, but it is worth here outlining a major design pattern used in IDO and the Ontology for General Medical Science (OGMS) [28], as it will emerge repeatedly below, namely, the relationship between **disorder**, **infectious disorder**, **infection**, **infectious disease**, and **infectious disease course**. Broadly speaking, a **disorder** is a clinically abnormal material part of an organism. **Infectious disorders** - the sort of **disorder** within the scope of IDO - are (1) **disorders** caused by some **infectious pathogen**, (2) a species of **infection** brought into existence by some **infectious pathogen**, and (3) the physical bases of **infectious diseases**. **Infectious diseases**, in turn, may manifest in an i**nfectious disease course**, such as in the exhibiting of various physical symptoms.

| Label | Definition |
|---|---|
| OGMS:**disorder** | Material entity that is a clinically abnormal part of an extended organism. |
| OGMS:**disease** | Disposition to undergo pathogenic processes that exists in an organism because of one or more disorders in that organism. |
| OGMS:**disease course** | Totality of all processes through which a given disease instance is realized. |
| IDO:**infectious disorder** | Disorder that is part of an extended organism which has an infectious pathogen part that exists as a result of a process of formation of disorder initiated by the infectious pathogen. |
| IDO:**appearance of disorder** | Process by which a disorder comes into existence. |
| IDO:**infection** | Material entity that is part of an extended organism that has some pathogen as part, which participates in the formation of the material entity by invading tissues of the organism. |
| IDO:**process of establishing an infection** | Process by which an infectious agent or infectious structure, established in a host, becomes part of an infection in the host. |
| IDO:**infectious disease** | Disease whose physical basis is an infectious disorder. |
| IDO:**opportunistic infectious disposition** | Infectious disposition to become part of a disorder only in organisms whose defenses are compromised. |
| IDO:**infectious disease course** | Disease course that is the realization of an infectious disease. |
| IDO:**infectious agent** | An organism that has an infectious disposition. |
| IDO:**pathogen** | A material entity with a pathogenic disposition. |
| IDO:**pathogen disposition** | Disposition borne by a material entity to establish localization |

6<... 

...

A Fourfold Pathogen Reference Ontology Suite

| | |
|---|---|
| | in, or produce toxins that can be transmitted to, an organism or acellular structure, either of which may form disorder in the entity or immunocompetent members of the entity's species. |
| IDO:**infectious disposition** | Pathogenic disposition borne by a pathogen to be transmitted to a host and then become part of an infection in that host or in immunocompetent members of the same species as the host. |
| IDO:**parasite** | An organism bearing a parasite role. |
| IDO:**parasite role** | A symbiont role borne by an organism in virtue of the fact that it derives a growth, survival, or fitness advantage from symbiosis while the other symbiont's growth, survival, or fitness is reduced. |
| IDO:**acellular structure** | Object consisting of an arrangement of interrelated acellular parts forming an acellular biological unit that is able to initiate replication of the structure in a host. |
| IDO:**infectious structure** | Acellular structure that bears an infectious disposition. |
| IDO:**virulence** | Quality that inheres in a pathogen and is the degree to which realizations of associated disease caused by the pathogen become severe or fatal. |
| IDO:**adhesion factor** | Biological macromolecule that has an adhesion disposition. |
| IDO:**adhesion disposition** | Disposition borne by a macromolecule that is the disposition to participate in an adhesion process. |
| IDO:**organism population** | Aggregate of organisms of the same species. |
| IDO:**antifungal disposition** | Disposition to kill or inhibit the development or reproduction of fungal organisms. |
| IDO:**resistance to drug** | A protective resistance that mitigates the damaging effects of a drug. |
| IDO:**protective resistance** | Disposition that inheres in an entity by virtue of it having some part which is disposed to mitigate damage to the entity. |
| MONDO:**mycosis** [29] | An infection that is caused by a fungus. |
| OPL:**parasite organism** [30] | An organism living in, with, or on another organism in parasitism. |
| OPL:**parasite lifecycle stage** | A life cycle stage of a parasite. |
| ChEBI:**drug** [31, 32] | Any substance which when absorbed into a living organism may modify one or more of its functions. |
| GO:**negative regulation of biological process** [25] | Any process that stops, prevents, or reduces the frequency, rate or extent of a biological process. |
| GO:**pathogenesis** | Process that is the realization of a pathogenic disposition inhering in a pathogen or pathogen population, having at least the proper process parts: (1) pathogen transmission, (2) establishment of localization in host, (3) process of establishing an infection, and (4) appearance of disorder. |
| FOODON:**fungus** [33] | Member of the group of eukaryotic organisms in the kingdom Fungi that includes unicellular microorganisms such as yeasts and molds, as well as multicellular fungi that produce familiar fruiting forms known as mushrooms. |
| MONDO:**cutaneous mycosis** [34] | Mycosis that involves the integument and its appendages, including hair and nails. |
| MONDO:**subcutaneous mycosis** | Mycosis that results in infection located in skin or located in subcutaneous tissue, which penetrate the dermis or even deeper during or after a skin trauma. |
| MONDO:**superficial mycosis** | Mycosis that is limited to the stratum corneum and essentially elicits no inflammation. |
| MONDO:**systemic mycosis** | Mycosis that involves the lungs, abdominal viscera, bones and or central nervous system. |

***Table 1****: Select Classes Reused from the Ontology for General Medical Science (OGMS), the Infectious Disease Ontology (IDO), the National Cancer Institute Thesaurus (NCI Thesaurus), the Ontology for Parasite Life Cycle (OPL), and the Chemical Entities of Biological Interest Ontology (ChEBI)*

Each ontology in the suite was rigorously evaluated to ensure scientific accuracy as well as ontological coherence. The inclusion of terminology was derived from extensive reviews of relevant, up-to-date authoritative literature, discussions with subject-matter matters from the relevant domain, and consensus-building exercises aimed at avoiding common modeling pitfalls, such as engagement in verbal disputes over labels. All aspects of ontology development, including addition of new terms, were driven by the needs of researchers investigating relevant domains. Consequently, no member of the suite is considered exhaustive but instead remains sensitive to evolving knowledge [1]. The Ontobee repository [35] and BioPortal [36] were used to investigate potentially relevant terms for each reference ontology.

**Results**

*The Virus Infectious Disease Ontology*

The Virus Infectious Disease Ontology (VIDO) serves as an extension of the Infectious Disease Ontology (IDO) Core, designed to represent the entities, processes, and relationships specific to viral infectious diseases. Its scope encompasses the detailed characterization of viruses, including their taxonomy, genetic composition, and replication mechanisms. It also addresses viral diseases, their clinical manifestations, and the biological processes involved in host-virus interactions. It consists of 36 unique classes and - leveraging existing resources where possible - reuses 400 existing ontology terms and 43 object properties from other ontologies. VIDO was developed based on virology literature [37, 38, 39, 40, 41]; subject-matter experts were consulted regarding definitions of key terms, such as the definitions for virus type terms that were minted for VIDO.

*A Fourfold Pathogen Reference Ontology Suite***Major Classes. Table 2** presents major VIDO classes. VIDO's central class is **virus** which is imported from the NCBITaxon and relabeled from "Viruses" to reflect the OBO convention that class names not be pluralized. Viruses are classified firstly in terms of their material structure, following the Baltimore Classification [37], where virus species are categorized into seven groups based on what steps members of a virus species must take during the viral replication cycle [38, 39, 40, 41]. For example, a positive-sense single-stranded RNA virus, such as SARS-CoV-2, has a genome which can be immediately translated into viral proteins upon entry into a cell, while a double-stranded DNA virus, such as *Orthopoxvirus variola*, must undergo transcription into messenger RNA before translation into viral proteins can proceed. VIDO reuses classes from GO reflecting the viral capsid composition as well as virus tropism [25], as well as terms from PRO to represent viral proteins [26], ChEBI for terms such as DNA and RNA [31, 32], and so on.

| VIDO Label | VIDO Definition |
|---|---|
| **virus** | Acellular self-replicating organic structure characterized by the presence of RNA or DNA genetic material and dependent on a host for replication |
| **virion** | Virus that is in its assembled state consisting of genomic material (DNA or RNA) surrounded by coating molecules |
| **viral infection** | Infection that has as part some virus |
| **viral infectious disorder** | Infectious disorder that has some virus as part, that exists as a result of a process of formation of disorder initiated by the virus |
| **viral infectious disease** | Infectious disease caused by a virus participating in a pathological process initiated by the virus within a host organism |
| **viral infectious disease course** | Infectious disease course that is the realization of a viral infectious disease |
| **viral pathogenesis** | Pathogenesis process realization of a pathogenic disposition inhering in a virus or virus population, having at least the proper process parts: (1) pathogen transmission, (2) establishment of localization in host, (3) process of establishing a viral infection, and (4) appearance of a virus disorder |
| **virus replication** | Biological process in which a virus synthesizes its genetic material and proteins within a host cell, resulting in the formation of new virions |
| **bacteriophage** | Virus which infects and replicates within or on bacteria or archaea |
| **double-stranded DNA virus** | Virus that has its genetic material encoded in double-stranded DNA and replicates using DNA polymerase |
| **positive-sense single-stranded RNA virus** | Virus with genetic material encoded in single-stranded RNA that can be translated directly into proteins |
| **virus replication** | Replication process in which a virus containing some portion of genetic material inherited from a parent virus is replicated |

| virus generative stage | Infectious structure generative stage that is a temporal subdivision of a virus developmental process |
|---|---|
| virus attachment stage | Virus generative stage during which a virion protein binds to molecules on the host surface or host cell surface projection |
| virus uncoating stage | Virus generative stage during which an incoming virus is disassembled in the host cell to release a replication-competent viral genome |
| negative regulation of virus attachment | Negative regulation of a virus replication process that stops, prevents, or reduces the frequency of some virus attachment stage |
| negative regulation of virus penetration | Negative regulation of coronavirus replication that stops, prevents, or reduces the frequency of some virus penetration stage |

*Table 2: Major VIDO Classes*

Though viruses are classified firstly in terms of their material structure, owing to their infectious nature, subclasses are secondarily classified as instances of the IDO class **infectious structure**, which are disposed to transmit to and become part of an **infection** in some host. This is captured in the OWL assertion that **infectious structure** is equivalent to:

**acellular structure** and **has disposition** some **infectious disposition**

The class **acellular structure** is a subclass of **material entity** – a broad BFO class that includes any entity that has some matter as part – instances of which do not have any parts that are cells. To illustrate **infectious disposition**: consider that SARS-CoV-2 virions are disposed to be transmitted to hosts via respiration, localize within host cells, leading to infection and disorder. In this respect, they are said to bear some **infectious disposition**. This highlights the fact that viruses are evolutionarily selected to infect hosts. Moreover, by adopting the distinction between **infectious structure** from **infectious disposition**, we distinguish the infectious potential of viruses from the underlying material basis that grounds that potential. Additionally, we disambiguate the **infectious disposition** of a given **virion** from its contribution to any associated **viral disease**, which in turn grounds the **viral disease course** that is the manifestation of disease in a host. More concretely, SARS-CoV-2 being infectious is in part explained by its being a positive-sense single stranded RNA virus; a host developing the disease COVID-19 is partially explained by SARS-CoV-2 being

*A Fourfold Pathogen Reference Ontology Suite*

infectious; a host manifesting various symptoms following infection is partially explained by their having COVID-19.

**Infectious disposition** provides a link to pathogens more generally, as it is a subclass of **pathogenic disposition** – borne by entities to localize in and form disorders within a host. Notice that an **infectious disposition** is understood as the disposition of the virus to be transmitted into the host and to initiate replication. This is not to be confused with an **infectious disease**, which is the disposition of an affected organism to undergo pathological processes due to a disorder. The class **pathogen** on the other hand is asserted to be equivalent to:

> **material entity** and **has disposition** some **pathogenic disposition**

From which it follows that **infectious structure** is a subclass of **pathogen**, as displayed in **Figure 1**.

Additionally, subclasses of **virus** such as **bacteriophage** – viruses which infect and replicate within or on bacteria – are equivalent to:

> **virus** and **has disposition** some **infectious disposition**

And since **virus** is a subclass of **acellular structure**, it follows that **bacteriophage** is an inferred subclass of **infectious structure** and **pathogen**, just as one would expect. These examples reflect the general design strategy underwriting VIDO: from the material structure and infectious dispositions of viruses, one can infer plausible alternative taxonomic virus classifications.

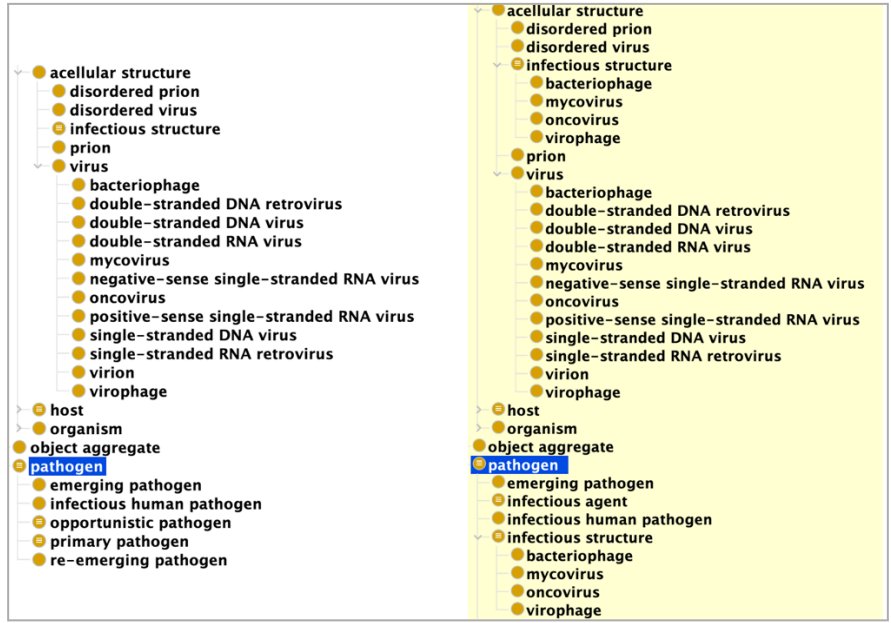

*Figure 2*: *Protégé Display of a Portion of Asserted and Inferred VIDO Hierarchies*

A core design pattern of VIDO is the representation of the virus replication cycle [38, 42], which encompasses the sequence of stages viruses propagate through to produce new virions within a host organism. The virus replication cycle covers crucial aspects of overall **viral pathogenesis**, understood as involving virus transmission, localization, establishment of infection, and the appearance of a viral infectious disorder. This cycle begins with the attachment of the virus to specific receptors on the host cell surface, followed by entry into the host cell through processes such as membrane fusion or endocytosis. Following penetration into a cell, a virion initiates replication, which varies considerably based on the type of virus, as characterized in the Baltimore Classification. Ultimately, virus replication must proceed to messenger RNA, which is in turn used to synthesize viral proteins, then assembled into virions and released from the host cell.

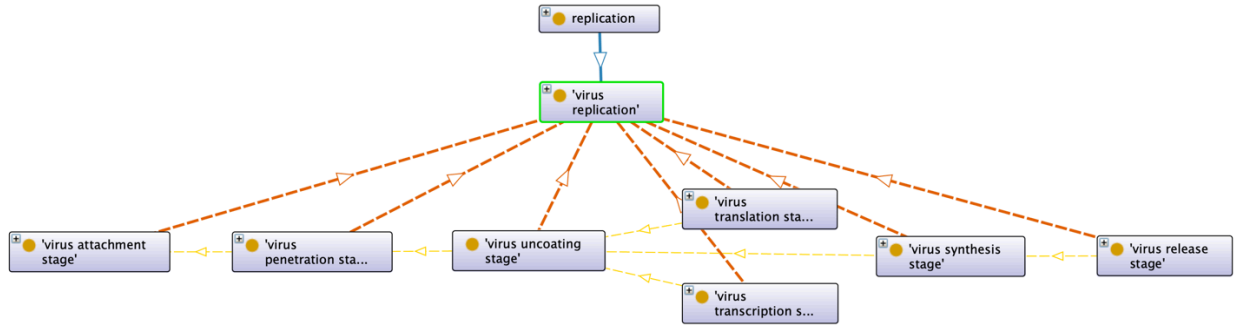

*Figure 3: Virus replication in VIDO*

**Figure 3** displays how the virus replication life cycle is represented in VIDO. Any instance of **virus replication stage** is necessarily part of some instance of **virus replication process**, as illustrated with the red arrows reflecting the **part of** relationship reused from RO. Similarly, any instance of **virus attachment stage** precedes any instance of the other stages, as shown by the yellow arrows reflecting the transitive **preceded by** relation. **Virus transcription** generally precedes **virus translation**, with the exception that positive-sense RNA virus genomes act directly as messenger RNA for immediate translation into viral proteins [43]. The Protégé editor from which the figure was generated relies on OWL for diagram generation, but OWL is not amenable to representing conditional scenarios. While SPARQL is useful for querying RDF, the Protégé SPARQL plugin does not support inserting inferred triples to the graph. For both, we leverage SWRL [44] rules where "^" reflects logical conjunction and "?" is prepended to variables:

```
virion(?x) ^ pssRNA(?x) ^ virusReplication(?p) ^ virusSynthesisStage(?y) ^ partOf(?y, ?p)
^ virusTranslationStage(?z) ^ partOf(?z, ?p) ^ participatesIn(?x, ?y) ^ participatesIn(?x,
?z) -> precededBy(?y, ?z)

virion(?x) ^ dsDNA(?x) ^ virusReplication(?p) ^ virusTranslationStage(?y) ^
partOf(?y, ?p) ^ virusTranscriptionStage(?z) ^ partOf(?z, ?p) ^ participatesIn(?x, ?y) ^
participatesIn(?x, ?z) -> precededBy(?y, ?z)
```

In words, if a **virion** that is an instance of the class **positive-sense single-stranded RNA** virus and **participates in** synthesis and translation stages of the same replication cycle, then the **preceded by** relation will be asserted to hold between the synthesis and translation stages. Similarly, if a **virion** that is an instance of the class **double-stranded DNA** virus and participates in translation and transcription stages of the same replication cycle, then the **preceded by** relation will be asserted to hold between the translation and transcription stages. In this way, VIDO respects the temporal orderings among stages in the virus replication cycle without falsely asserting that **virus transcription** in every case precedes **virus translation**.

Unambiguous representation of the virus replication has clear value for development of intervention and treatment strategies in the domain of rational drug design. For example, only host cells with specific features are susceptible to attachment by SARS-CoV-2 [3]. In humans, the standard route for successful infection involves virion attachment to host alveolar epithelial cells through angiotensin-converting enzyme 2 (ACE2) receptors [45, 46], defined in the PRO [26]. These cells may be characterized as bearing an **adhesion disposition**, defined in IDO as macromolecules disposed to participate in adhesion processes. Following the virus replication cycle, cell penetration often follows attachment, where host cell cleavage is aided by trans-membrane protease serine 2 (TMPRSS2), before SARS-CoV-2 cell membrane fusion [47]. An ontological characterization of the SARS-CoV-2 replication cycle allows for inferences about possible targets for drugs and interventions designed to disrupt that cycle. Such disruptions can, moreover, be represented ontologically by way of extensions of the GO class **negative regulation of biological process** - any process that stops, prevents, or reduces the frequency, rate or extent of a biological process - such as **negative regulation of viral life cycle**. Extended to virus replication stages, we generate **negative regulation of virus attachment**, and so on for each stage.

**Extensions.** The modular design of VIDO supports its extension into more specific virus domains, ensuring its continued relevance as new challenges in virology emerge. The Coronavirus Infectious Disease Ontology (CIDO) is perhaps the most widely-used extension, having been leveraged in the search for coronavirus interventions, treatment exploration, and basic research [5]. Since the development of VIDO, the VIDO and CIDO teams have worked to bring the latter under the guardrails of the former [16]. Additional extensions of VIDO have emerged since its development, such as the IoT-MIDO project [48] which leverages infectious disease vocabularies based on IDO to explore patient monitoring and risk assessment, clinical management of patients with infectious diseases, as well as epidemic risk analysis and surveillance. From another direction, the Covid19-IBO project aims to integrate virus-specific vocabularies from several ontologies containing terms related to COVID-19 in the interest of gaining insights into the impact this disease had on the banking sector in India; the authors provide a schema to identify similarities and differences across these virus ontologies [49].

Looking ahead, there are several virus-specific extensions of IDO which are within scope of alignment to VIDO. Specifically, the HIV Ontology (IDOHIV) [12], the Influenza Ontology (IDOFLU) [7], and the Dengue Fever Ontology (IDODEN) [13] - none of which have been updated since 2017 [3, 11] - should be brought into alignment with VIDO. Doing so will have the benefit of aligning each to the updated IDO, promoting semantic interoperability among IDO extensions, and encouraging further development of virus-specific ontologies within this ecosystem.

*The Bacteria Infectious Disease Ontology*

The Bacteria Infectious Disease Ontology (BIDO) extension of IDO is designed to serve as a reference ontology for bacterial pathogens and bacterial pathogenesis more generally. Its scope encompasses the detailed characterization of bacteria, including their taxonomy, genetic composition, and reproduction mechanisms. It also addresses bacterial diseases, their clinical manifestations, and the biological processes involved in host-bacteria interactions. It consists of 37 unique classes and - leveraging existing resources where possible - reuses 1400 existing ontology terms and 55 object properties from other ontology projects. BIDO was developed based on bacteria literature [53, 54, 55, 56]; subject-matter experts were consulted regarding definitions of key terms, such as the definitions for bacteria type terms that were newly created for BIDO.

**Major Classes and Relations.** BIDO introduces - as subclasses of the undefined NCIBTaxon term **bacterium** (relabeled from "Bacteria" to reflect the OBO convention that classes have singular names) - new terms for a variety of bacterium types. **Figure 4** provides example subclasses, such as **bacilli bacterium**, **spirilla bacterium**, **cocci bacterium**, and **spirochetes bacterium**, which reflect bacteria classifications determined by cellular morphology.

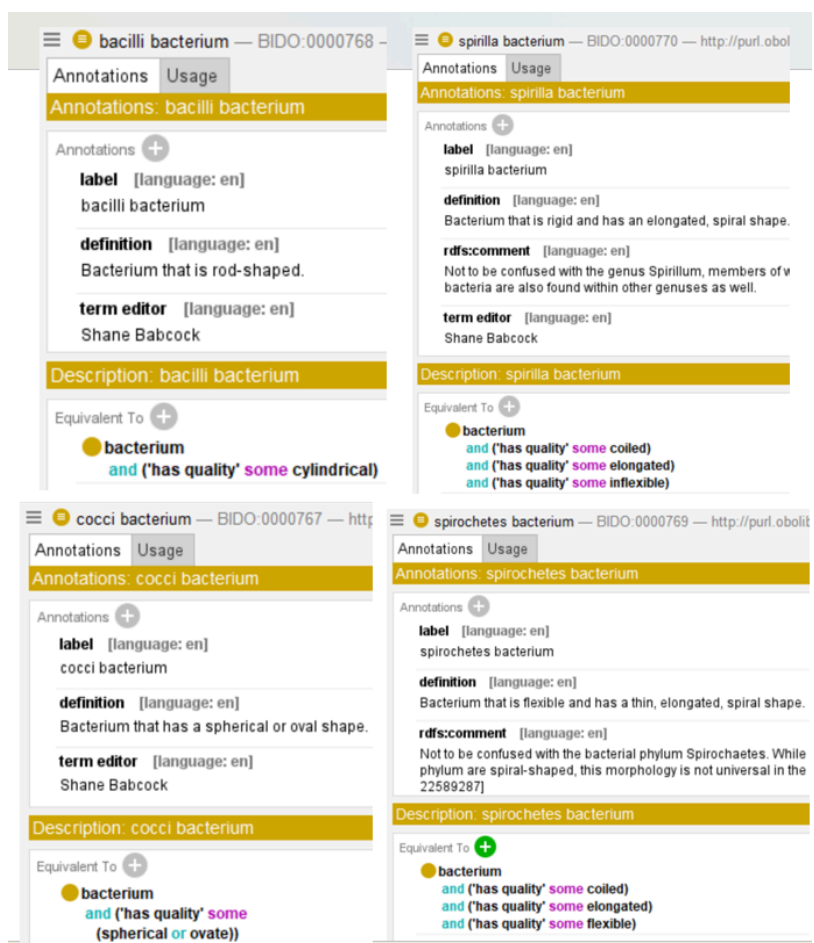

*Figure 4: Bacteria Type Vocabulary in BIDO*

BIDO provides rich semantics around these terms by importing and additional terms broadly relevant to the bacterial domain from several OBO Foundry ontologies, such as IDO, ChEBI, PRO, and GO. From ChEBI, molecular entity terms are imported that factor in bacterial

*A Fourfold Pathogen Reference Ontology Suite*pathogenesis, such as **lipopolysaccharide**, **capsular polysaccharide**, and **bacteriocin**. PRO provided a source for various bacterial protein terms while GO imports included bacterial cell components such as **Gram-positive-bacterium-type cell wall**, **pilus**, **capsule**, **slime layer**, and so on, as well as important, cellular process terms such as **entry of bacterium into host cell**, **aerobic respiration**, **fermentation**, and spore dispersal, among others. Additionally, the Ontology of Microbial Phenotypes (OMP) [57] provided phenotype terms characteristic of different bacterium types, such as **obligate anaerobe**, **obligate aerobe**, **facultative anaerobe**, and so on, while the Clinical Measurement Ontology (CMO) [58] provided measurement data terms relating to bacterial infection, such as **bacterial infection severity measurement**, **bacterium count**, and **bacterial infection severity score**. As illustrated in **Figure 5**, BIDO includes various logical axioms connecting bacterial pathogenicity to key such imported terms, as well as terms from IDO such as **toxin**, **virulence factor**, **adhesion factor**, and **invasion factor**. Imported terms are used to provide logical definitions for various bacterium types such as **aerobic bacterium, anaerobic bacterium, spirilla bacterium**, discussed in more detail below.

*Figure 5*: *Semantic Enrichment of Existing Bacteria Vocabulary in BIDO*

BIDO's major classes extend directly from terms in either IDO or OGMS, following the "hub-and-spoke" methodology employed in VIDO development [16]. BIDO key terms and definitions are displayed in **Table 3**.

| BIDO Label | BIDO Definition |
|---|---|
| **spirilla bacterium** | Bacterium that is rigid and has an elongated, spiral shape. |
| **spirochetes bacterium** | Bacterium that is flexible, and has a thin, elongated, spiral shape. |
| **bacterial infection** | Infection that has as part some bacterium. |
| **bacterial toxin disorder** | Disorder the formation of which involves some bacterial toxin. |
| **bacterial infectious disorder** | Infectious disorder that has some bacterium as part, that exists as a result of a process of formation of disorder initiated by the bacterium. |
| **bacterial infectious disease** | Infectious disease that has its material basis in some bacterial infectious disorder. |
| **bacterial infectious disease course** | Infectious disease course that is the realization of a bacterial infectious disease. |
| **bacterial pathogenesis** | Pathogenesis process realization of a pathogenic disposition inhering in some bacterium or bacteria population, having at least the proper process parts: (1) toxin biosynthetic process, and (2) appearance of disorder. |
| **bacterial adhesion disposition** | Adhesion disposition borne by a macromolecule that is part of some bacterium that is the disposition to participate in the adhesion of the bacterium to a host. |
| **bacterial adhesion factor** | Adhesion factor that is part of some bacterium. |
| **bacteria population** | Organism population whose members are bacteria. |
| **bacteria colony** | Colony whose members are bacteria. |
| **appearance of bacterial infectious disorder** | Appearance of disorder by which a bacterial infectious disorder comes into existence. |
| **process of establishing a bacterial infection** | Process of establishing an infection in which a bacterium participates. |
| **bacterial pathogenesis involving infection** | Bacterial pathogenesis having at least the proper process parts: (1) establishment of localization in host, (2) process of establishing a bacterial infection, and (3) appearance of bacterial infectious disorder. |

*Table 3: Major BIDO Classes*

BIDO distinguishes between **bacterial toxin disorder** and **bacterial infectious disorder**. Instances of the latter are in every case instances of **bacterial infection** and serve as the material basis for **bacterial infectious disease**. In contrast, the class **bacterial toxin disorder** covers any disorder involving bacterial toxins, whether associated with an infection by the relevant toxin-producing bacteria or not.[3] An example to illustrate this distinction concerns the causative agent in **food botulism**, *Clostridium botulinum*, a bacterial pathogen that is not infectious. When a person consumes food contaminated by toxins produced by *C. botulinum*, the ingested toxins often cause disorder in that person, leading to the disease. The disorder is not, however, a **bacterial infectious disorder**, since the bacterium is not disposed to invade or be transmitted to other potential hosts [3, 16]. Of course, some bacterial toxin disorders *are* instances of **infectious disorder** as when toxins secreted by bacteria serve as virulence factors aiding infection of a host by, say, helping bacteria bypass innate and adaptive immune responses.

---
[3] This distinction parallels that drawn in IDO between **infectious disorder** and **infection** [12].

Similarly, some pathogenic bacteria may be infectious while others are not. For example, *Yersinia pestis* and *Vibrio cholerae* are disposed to be transmitted and become part of some infection and so be counted as **infectious agents** in the creation of **bacterial infectious disorders**. Even so, bacteria need not be infectious to count as pathogenic; they need only exhibit **virulence**. That is, any instance of a **bacterium** which bears a **virulence** quality counts as an instance of a **pathogen**. As pathogens, such bacteria bear instances of **pathogenic disposition**, and are consequently disposed to establish localization in or produce toxins transmittable to a host to form disorder.

Modeled on VIDO's **viral pathogenesis**, BIDO introduces **bacterial pathogenesis**, which has the subclass axiom:

**has part** some **toxin biosynthetic process** and (**has participant** some **bacterium**)

This axiom leverages the GO term **toxin biosynthetic process**, characterizing the formation of toxins by cells or organisms, including bacteria that cause disease in various organisms. Unlike **viral pathogenesis***,* however, we cannot assert that instances of **bacterial pathogenesis** in every case involve an **establishment of localization in host** or a **process of establishing an infection**, as a part. As discussed above there are examples of bacterial pathogenesis in which bacterial toxins cause disease in a host but are produced by bacteria that never localize in the diseased organism. Additionally, BIDO allows that **bacterial pathogenesis** may involve infections caused by opportunistic pathogens that do not engage in any **pathogen transmission process**, such as when bacteria colonizing the surface of a host's skin realize instances of **infectious disposition** when that protective barrier is ruptured.

BIDO includes newly added axioms, connecting various imported terms to relevant IDO terms. To illustrate, consider the GO term **pilus**. Pili are the hair-like appendages found on many bacteria which help facilitate the adhesion of pathogenic strains of bacteria to bodily tissues. Pili thereby increase bacteria replication rates, allow bacteria to colonize host cells, and facilitate tissue infection. Pili thus contribute to the virulence of many pathogenic bacteria. BIDO represents this complexity around the term **pilus** with the following equivalency axiom:

> **cell projection** and **part of** some **bacterium** and **has role** some **antigen role** and **has disposition** some **bacterial adhesion disposition** and **has disposition** some **virulence factor disposition**

As logically defined, **pilus** is an inferred subclass of IDO's **adhesion factor** and **virulence factor**.

Other examples of bacterial virulence factors are represented by terms imported from ChEBI. A **capsular polysaccharide** is a polysaccharide capsule found on the cell surface of many bacteria that enables both their adhesion to surfaces and their evasion of host immune responses, as well as providing protection from toxins. **Lipid A** is the lipopolysaccharide portion of the outer membrane of gram-negative bacteria responsible for its endotoxicity. As with **pilus**, we assert that both **capsular polysaccharide** and **lipid A** have a **virulence factor disposition**, so both are inferred subclasses of IDO's **virulence factor**. In the case of **lipid A**, instances have some **endotoxin disposition**, and so it is an inferred subclass of **endotoxin**. As development of BIDO continues, we will add similar axioms to other imported terms from the GO, PRO, and ChEBI.

**Extensions.** There are currently four existing bacterial infectious disease ontologies extending from IDO which would stand to benefit from refactoring as domain ontology extensions of BIDO:

1. Staphylococcus aureus Infectious Disease Ontology (IDO-Staph) [59]
2. The Brucellosis Infectious Disease Ontology (IDOBRU) [60]

3. The Meningitis Infectious Disease Ontology (IDOMEN) [61]
4. The Bacterial Clinical Infectious Disease Ontology (BCIDO) [62, 63]

Each requires updating to the most recent versions of IDO and BFO, during which refactoring to support BIDO as a common reference ontology can be conducted. This seems particularly impactful for BCIDO, designed to represent within its scope bacterial infections, bacteria, and antibiotic treatment, excluding those associated with *Mycobacteria*. Presentations of BCIDO [63] suggest there are many ontology terms that would be included in a bacterial infectious disease reference ontology but also suggest that BCIDO is not particularly modular. We maintain that it would thus be in the interest of modularity and reuse, that BIDO and BCIDO communities work towards a division of labor in this domain.

Given the scope of BIDO as a reference ontology, additional potential extensions cover the broad domain of bacterial infectious diseases, and may include ontologies for tuberculosis, infective endocarditis, chlamydia, *E. coli* infection, and so on.

## The Mycosis Infectious Disease Ontology

The Mycosis Infectious Disease Ontology (MIDO) is an open-source biomedical ontology built with the purpose of providing standardized ontological representations specific to fungal infectious diseases. Its scope encompasses detailed characterization of fungi, including their taxonomy, genetic composition, and reproduction mechanisms. It also addresses fungal diseases, their clinical manifestations, and the processes involved in host-fungus interactions. MIDO consists of 71 unique classes and - leveraging existing resources where possible - reuses 526 existing ontology terms and 39 object properties from other ontology projects. MIDO aims to bring fungal infectious diseases in conversation with other OBO Foundry ontologies related to infectious diseases. It was developed based on fungi literature [44, 65] and engagement with subject-matter experts in the UNITE community [66] which maintains an open-source database for fungal taxonomy and nomenclature.

*A Fourfold Pathogen Reference Ontology Suite*

**Major Classes and Relations.** Fungi are a diverse set of eukaryotic organisms that play an important role in our ecosystem. **Table 4** highlights key classes and definitions introduced by MIDO, which extends the FOODON [33] term **fungus** by introducing subclasses for **yeast**, **mold**, and **dimorphic fungus**. Moreover, terms for the anatomical structure of fungi are imported from the OBO Foundry Fungal Gross Anatomy Ontology (FAO) [67], such as the general class **fungal structure**, and subclasses such as **hyphal tip**, **sterigma**, and **mycorrhiza**.

Certain fungus types can cause hosts to develop fungal infections - captured by the MONDO Disease Ontology (MONDO) [61] term **mycosis** - which vary in severity owing to the host tissues they affect. MIDO reuses other terms from MONDO relevant to various mycoses, such as the commonly drawn fourfold distinction between **superficial mycosis**, **cutaneous mycosis**, **subcutaneous mycosis**, and **systemic mycosis**. An example of **systemic mycosis** occurs through the inhalation of fungal spores, such as those produced by *Histoplasma capsulatum*. The MONDO term **histoplasmosis**[4] is reused here to represent the associated MIDO **fungal infectious disease**. More specifically, inhalation of *Histoplasma capsulatum* spores above a certain threshold may result in the formation of **fungal infectious disorder** in a host, which serves as the physical basis for **histoplasmosis**. This **fungal infectious disease** may then be realized in a **histoplasmosis infectious disease course**, often involving a variety of symptoms such as chills, fevers, headaches, and so on. This basic design pattern inherited from IDO is generalizable to other fungal infectious diseases, such as when an instance of **aspergillus** has colonized and grows within a host's lungs or sinuses, forming an **aspergilloma** - a **fungal infectious disorder** - which is the physical basis for MONDO's **aspergillosis** - a **fungal infectious disease** - that may be realized in an **aspergillosis infectious disease course**.

---

[4] MONDO cross-lists this definition with NCI, but it is worth noting the Human Phenotype Ontology (HPO) also contains a term **histoplasmosis** defined as, roughly, an opportunistic fungal infection caused by *Histoplasma capsulatum*. MONDO's classification fits most closely with the OGMS's treatment of **disease**, which we adopt in MIDO, hence the use of MONDO's terms rather than those found in HPO. That said, in the interest of interoperability, we intend to construct mappings between these ontologies, to identify where they share and differ in semantics.

*A Fourfold Pathogen Reference Ontology Suite*

| MIDO Label | MIDO Definition |
|---|---|
| **yeast** | Fungus that is unicellular through its development cycle. |
| **mold** | Fungus that is multicellular through its development cycle, exhibiting mycelium networks. |
| **dimorphic fungus** | Fungus is unicellular or multicellular through its development cycle depending on environmental conditions. |
| **fungal infectious disorder** | Infectious disorder that has some fungus as part, that exists as a result of a process of formation of disorder initiated by the fungus. |
| **aspergilloma** | Fungal disorder which contains as part aspergillus localized in the lungs or sinuses of a host. |
| **fungal infection** | Infection that has as part some fungus. |
| **fungal infectious disease** | Infectious disease caused by a fungus participating in a pathological process initiated by the fungus within a host organism. |
| **fungal infectious disease course** | Infectious disease whose physical basis is a fungal disorder that is clinically abnormal in virtue of the presence of the relevant fungus population. |
| **fungal pathogenesis** | Pathogenesis process realization of a pathogenic disposition inhering in a fungus or fungus population, having at least the proper process parts: (1) establishment of localization in or on host, (2) process of establishing a fungal infection, and (3) appearance of a fungal infectious disorder. |
| **fungal infectious agent** | Fungus that has an infectious disposition. |
| **opportunistic fungal pathogen** | Fungus bearing an opportunistic infectious disposition. |
| **azole** | Antifungal that targets fungal cytochrome P450 enzyme-lanosterol 14α-demethylase which converts lanosterol to ergosterol. |
| **polyene** | Antifungal that directly binds to and removes the ergosterol present in the fungal cell membrane. |
| **antifungal resistance** | Resistance to drugs that mitigates the damaging effects of antifungal treatments. |

*Table 4: Major MIDO Classes*

Following IDO, MIDO adopts that "pathogen" should be understood as indexed to specific species and maturity of potential hosts. For example, *Ceratocystis paradoxa* is a fungal plant pathogen responsible for significant annual loss of pineapple harvests [68], but which is not considered pathogenic to humans. This host- and context-sensitivity also underlies our treatment of opportunistic pathogens, which is particularly important for MIDO since many fungal infections are opportunistic. Following IDO, we maintain that opportunistic pathogens are not simply microbes that become pathogenic by virtue of an "opportunity"; rather, they are microbes that already bear a **pathogenic disposition**, which may remain latent under normal conditions and realized in others. This is reflected in **opportunistic fungal pathogen**, which is constrained by the logical axiom:

> **inheres in** some **fungus** and (**realized in** only (**process** and **has part** some **establishment of localization in host** and **has part** some **transmission process** and **has part** some **appearance of disorder** and not **occurs in** some **immunocompetent organism**))

This axiom indicates that any instance of **opportunistic fungal pathogen** is disposed to establish localization in a host that is not immunocompetent, following a transmission process, from which a disorder emerges. A common example of an **opportunistic fungal pathogen** is *Candida albicans*, a **yeast** that is commonly found in the human gut flora and often associated with thrush and gastrointestinal infections in immunocompromised human hosts. The alignment of fungal pathogen terminology with that of other infectious diseases, thus allows for exploration of secondary fungal infections. Of value is the identification of patients exhibiting weakened immune systems who may be thereby susceptible to fungal infections.

From another angle, fungal spores can provoke pathogenic responses that may, in clinical presentation, be difficult to distinguish from allergic reactions such as those induced by pollen inhalation. MIDO can serve as a framework for semantically distinguishing allergic responses from infectious pathogenic responses. Following IDO, allergic responses may be modeled as realizations of hypersensitivity dispositions involving **material entities** that bear allergen **roles** which participate in some **process** of hypersensitivity. In contrast, pathogen responses are modeled as **realizations** of **infectious dispositions** borne by **pathogens** that initiate the formation of **disorders** which are in turn the physical bases of **diseases**. When encoded in OWL, associated reasoners can distinguish between allergic and infectious etiologies for otherwise similar manifestations.

Diagnosing fungal infections is inherently challenging due to the diverse and complex nature of fungi [69]. Compounding these diagnostic challenges is an even more pressing issue: the growing resistance of fungi to antifungal treatments. Despite the identification of antifungal drugs that seem to maintain efficacy after decades of use as treatments, such as Amphotericin B instances of the MIDO class **polyene**, many antifungal drugs, including those represented in MIDO by the **azole** class, have shown decreasing efficacy worldwide due to the emergence of **antifungal resistance** inhering in fungal strains, such as *Aspergillus fumigatus* [70, 71]. MIDO offers an opportunity to address such challenges by providing well-defined representations of the mechanisms of antifungal resistance and the genetic factors contributing to it. Such enhancements may facilitate the study of resistance patterns, aiding the design of more effective treatment plans and the development of next-generation antifungal drugs [72]. Additionally, integrating MIDO with drug discovery databases will enable the identification of novel antifungal compounds. By connecting fungal genes and pathways with existing drug targets, researchers can leverage these links to develop new therapeutic options with greater precision and efficiency [73].

**Extensions.** Given how recent initiation of MIDO has been, there are currently no ontologies extending from it. Moreover, given pathogen research heavily favoring bacterial and viral pathogens, few OBO ontologies contain fungal infection classes or relations. Nevertheless, there are numerous reasons to take such modeling seriously, not the least of which owes to the considerable amount of data being curated by organizations such as UNITE, which run the risk of creating data silos. MIDO developers envision leveraging this ontology to provide a semantic layer connecting data from the UNITE community to other datasets concerning fungi.

Additionally, MIDO developers envision using MIDO to model incidence and spread of fungal infections, including rare and emerging mycoses, in the interest of supporting outbreak monitoring and response. Ontological characterizations of the complex interrelations of fungal pathogens and their hosts will, moreover, clarify understanding of fungal pathogenesis, perhaps opening new therapeutic research options and infection prevention strategies. MIDO is, however, less developed than either VIDO or BIDO, and so much work remains before it is able to make good on such potential. With that in mind, **Table 5** displays critical MIDO updates needed for future ontology development.

| Ontology Gap | Description | Potential Ontology Content | Potential Reuse |
|---|---|---|---|
| **Environmental and Ecological Context** | Represent fungi's roles in ecosystems to capture interactions and ecological significance. | **decomposer, symbiotic fungus, has ecological role** | Environment Ontology (ENVO) [74] |
| **Epidemiological Modeling** | Incorporate epidemiological data for fungal infection patterns, outbreaks, and transmission dynamics to help in forecasting and monitoring. | **fungal outbreak, fungal infection incidence, fungal infection incidence rate, fungal infection susceptibility** | APOLLO-SV [75]; MONDO |
| **Novel Therapeutic Approaches** | Represent new antifungal treatments, alternative therapies, and mechanisms to counter antifungal resistance. | **fungal vaccine, antifungal nanoparticle, chitin synthesis inhibitor** | Vaccine Ontology (VO) [76, 77]; FAO |

***Table 5**: Planned Updates to MIDO*

## *The Parasite Infectious Disease Ontology*

The Parasite Infectious Disease Ontology (PIDO) extension of IDO is designed to serve as a reference ontology for eukaryotic pathogens and parasitic pathogenesis more generally. Its scope encompasses the detailed characterization of parasites, including their taxonomy, genetic composition, and reproduction mechanisms. It also addresses parasite diseases, their clinical manifestations, and the biological processes involved in host-parasite interactions. PIDO consists of 13 unique classes and - leveraging existing resources where possible - reuses 1015 existing ontology terms and 44 object properties from other ontology projects. PIDO was developed based on contemporary parasite literature [78, 79, 80, 81, 82]; subject-matter experts were, as always, consulted regarding definitions of key terms and relations.

**Major Classes and Relations.** Historically the term "parasite" refers to pathogens that are neither bacteria nor viruses [78]. In part to avoid classifying, say, prions as parasites, IDO maintains parasites are instances of **organism** that bear some **parasite role**, which is itself a subclass of **symbiont host role**. Instances of the latter are borne by some organism in virtue of the fact that the organism derives growth, survival, or some fitness advantage from symbiosis, while its host's growth, survival, or fitness is reduced. Such a broad ontological characterization of parasites seems warranted, given their sheer diversity. Human parasites include organisms ranging from helminths (e.g. tapeworms and roundworms), protozoans (e.g. amoeba and ciliates), as well as ectoparasites described as parasites that live on the exterior of its host (e.g. fleas and lice).

The primary source for imported terms to PIDO is the Ontology for Parasite Lifecycle (OPL) [79]. OPL distinguishes between a **parasite life stage** and a **parasite organism**. The former is a **spatiotemporal region** – a class in BFO that is the spatial and temporal extent of some **process** – that encompasses some part of the life cycle of an organism, which is intended to cover the temporal extent over which a parasite is alive. This class is asserted equivalent to:

**life cycle stage** and **in taxon** some **parasite organism**

Where **life cycle stage** is imported from UBERON [83], and **in taxon** is imported from RO, meaning $x$ is **in taxon** $y$ just in case $y$ is an organism and either $x$ is part of, developmentally preceded by, derives from, secreted by, or expressed $y$. Building on these imports, PIDO introduces **parasite generative stage** as representing the processes which occur in the parasite life stage spatiotemporal region, that are the proper targets of rational drug interventions and treatments. Additionally, terms are imported by OPL from the Cell Line Ontology (CLO) to specify which tissues are targeted by parasites [84] during parasitic infections.

In many cases, parasites exhibit distinct phenotypes at different life stages. Some examples of terms reused for such representation include **plasmodium falciparum gametocyte**, **plasmodium falciparum gametocyte stage** and **plasmodium falciparum trophozoite stage**. Terms representing parasite diseases were reused from the extensive disease representation found in MONDO, where appropriate terms existed. Finally, when these were available, we imported terms representing existing drugs and vaccines that target parasites from ChEBI or VO [85]. Examples include **malaria, leishmaniasis, toxoplasmosis, chloroquine**, and **parasite vaccine**. The initial parasites represented within PIDO were chosen to reflect those that have vaccines within VO or the associated Vaccine Investigative Online Information Network (VIOLIN) database [86]. To represent the parasite life stages for the 23 parasitic eukaryotes listed in VIOLIN, we created new terms by searching for information on parasite life stages and species through a literature review on PubMed.

| Label | Definition |
|---|---|
| **parasite infection** | Infection that has as part some parasite. |
| **parasite infectious disorder** | Infectious disorder that exists as a result of a process of formation of disorder initiated by a parasite or parasite population. |
| **parasite infectious disease** | Infectious disease that has material basis in some parasite disorder. |
| **parasite infectious disease course** | Infectious disease whose physical basis is a parasite disorder that is clinically abnormal in virtue of the presence of the relevant parasite population. |
| **parasite pathogenesis** | Pathogenesis process realization of a pathogenic disposition inhering in a parasite or parasite population, having at least the proper process parts: (1) pathogen transmission, (2) establishment of localization in host, (3) process of establishing a parasite infection, and (4) appearance of a parasite infectious disorder. |
| **ectoparasite** | Parasite bearing an ectoparasite role. |
| **ectoparasite role** | Parasite role whose realization involves attachment to and interaction with the surface of a host. |
| **endoparasite** | Parasite bearing an endoparasite role. |
| **endoparasite role** | Parasite role whose realization involves location within and interaction with the body of a host. |
| **parasite infectious agent** | Parasite that has an infectious disposition. |
| **parasite generative stage** | Generative stage that is a temporal subdivision of a parasite developmental process. |
| **infection-causing parasite life stage** | Parasite life stage that initiates the formation of an infection within the host. |
| **disease-causing parasite life stage** | Parasite organism life stage that initiates the formation of a disease within the host. |
| **antiparasitic disposition** | Disposition to kill or inhibit the development or reproduction of parasites. |

*Table 6*: *Major PIDO Classes*

We can divide the terms in PIDO into two main groups. The first focuses on classifying parasite types. The Center for Diseases Control and Prevention (CDC) provides a list of parasitic infectious diseases against humans and domesticated animals that covers over 114 different species from 12 different clades. However, within the context of infectious disease, the focus is moved to members of three major polyphyletic categories; parasitic prokaryotes (internal parasite eukaryotic microbe), parasitic helminths (internal parasitic animals), and parasitic arthropods (ectoparasite animals). Instances of **parasite role** are specified by the location in the host where the parasite role is realized. For example, **ectoparasite** specifies that the accompanying **parasite role** is realized when located on a host's surface. The **endoparasite role** on the other hand is much broader as it covers instances of **parasite** that exist within a host's body, such as lumen, tissues, or internal organs. Core terms from PIDO are presented in **Table 6**.

The second major division of PIDO terms represent specific life stages within the context of a disease. Instances of **parasite life stage** are unique insofar as the given **parasite** can exhibit significant phenotypic differences between the **parasite infection** and disease states to evade the host immune system. Bilharzia, for example, is a disease caused by the *Schistosoma* flatworm. *Schistosoma* migrate and infect human hosts as S*chistosomia cercariae* before maturing as adults. However, bilharzia is caused by the **schistosomiasis gametes** laid by adult *Schistosoma* [88]. To represent these differences, a distinction is drawn in PIDO between **infection-causing parasite** and **disease-causing parasite**, represented as different life cycle stage types. A similar pattern can also be observed with prokaryotic parasites, such as the malaria parasite *Plasmodium falciparum*.

**Figure 6** shows a general design pattern for PIDO. Using malaria as an example, axiomatization of classes representing parasites, their participation as causative agents in disease, and associated treatment options, follow the pattern: parasite participates in some generative stage which is part of some parasite development process, which is negatively regulated by the realization of some antiparasitic disposition. For example, *Plasmodium falciparum* participates in some generative and developmental stages associated with its lifecycle, which sometimes lead to hosts exhibiting symptoms of malaria; malaria drugs such as chloroquine – which increases the levels of heme in host blood to toxic levels for the parasite - are then used to regulate the parasite developmental cycle.

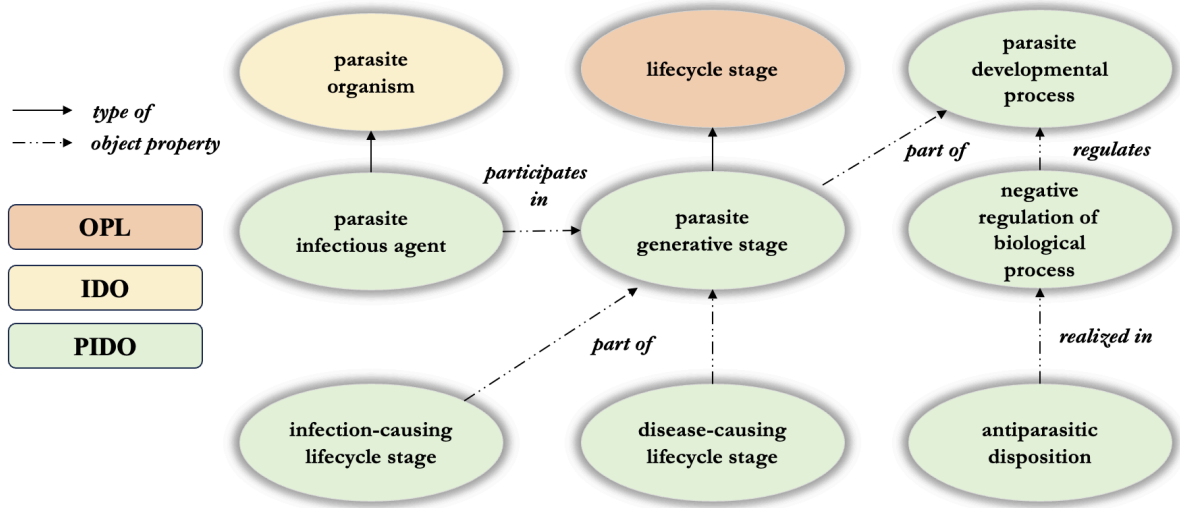

***Figure 6****: PIDO design pattern where "type of" relates instances and classes and object properties relate instances*

Drugs and vaccines that target parasitic protozoans, such a *Plasmodium falciparum,* are efficacious only against parasites within specific life cycles. Using a malaria vaccine as an example, we can represent within PIDO that efficacy against *P. falciparum*:

**RTS,S/AS01 immunizes against microbe** some **P. falciparum erythrocyte**

Logically connecting the vaccine to immunity provided against malaria in humans. Similarly:

**PFS25/28 in Matrix M immunizes against** some **P. falciparum gametocyte**

Is an axiom that logically characterizes a malaria vaccine that inhibits transmission of a parasite.

**Extensions.** There are currently two existing parasite infectious disease ontologies extending from IDO which would stand to benefit from refactoring as domain ontology extensions of PIDO:

1. Malaria Infectious Disease Ontology (IDOMAL) [89]
2. The Schistosomiasis Ontology (IDOSCHISTO) [90]

Each requires updating to the most recent versions of IDO and BFO, during which refactoring to support PIDO as a common reference ontology can be conducted. Developers of IDOMAL were challenged by the question of how best to model host, pathogen, and vector succinctly within an ontology. PIDO resolves this issue by leveraging **parasite life stages** in which a **pathogen** is involved, either with a host or as part of a vector. Moreover, IDOSCHISTO's simpler structure facilitates the mapping of life stage cycles into the appropriate PIDO format, illustrated in **Figure 7.**

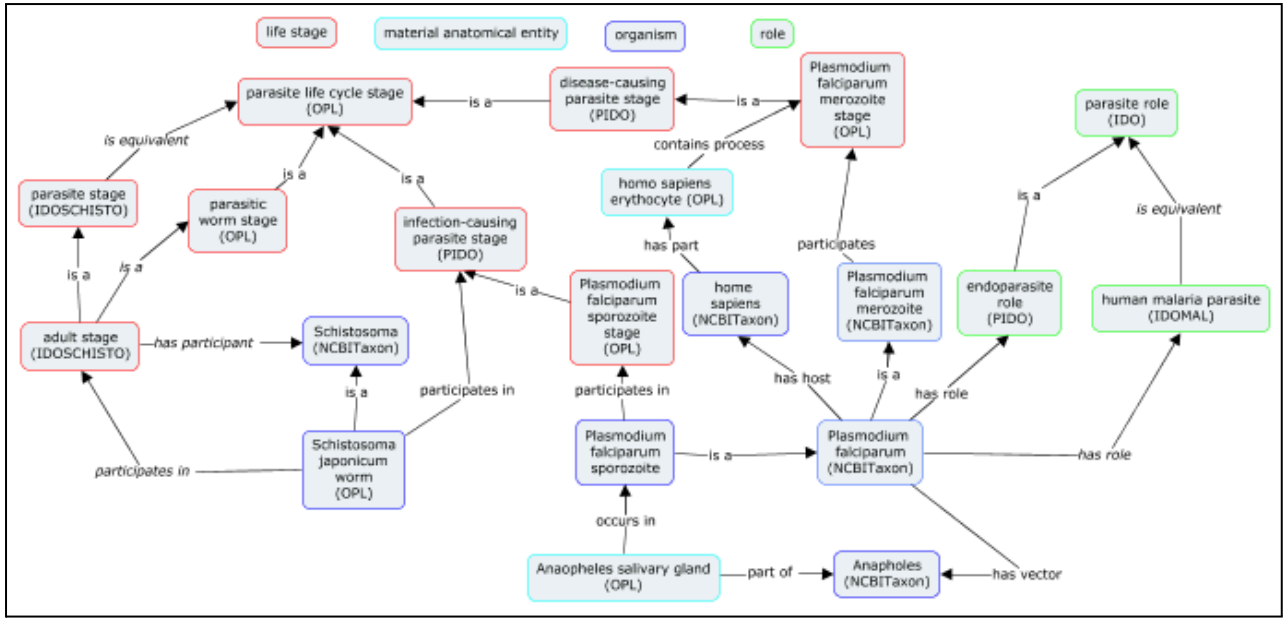

**Figure 7.** *Refactorization of IDOMAL and IDOSCHISTO using PIDO. New classes in PIDO enable construction of a unifying data model, where relations used for construction are shown in italics.*

Further extensions of PIDO will be focused on adding information to represent other infectious parasites. First, future endeavors will involve clarifying the internal division of PIDO, possibly splitting its scope between animal parasites and protozoan parasites. Second, we plan to develop PIDO so that it is able to represent more complex use cases. One such example would include scenarios where a parasite functions as a vector for another pathogen, like a tick ectoparasite being a vector for the bacteria that cause Lyme disease. Extending PIDO would involve adding new classes to ontologically represent the relation between a **parasite**, a **parasitic host**, and a **parasitic vector**. The parasitic vector is a complex type of entity to represent ontologically, as the life cycles of instances would presumably be associated with an additional symbiont stage where the parasite may co-exist in a non-parasitic relationship, such as malaria parasites within mosquitoes. Creating an ontological representation of such a scenario would allow for the semantic modeling of the tripartite relationship, and thus potentially for the discovery of new relations implicit in data collected on various stages of development of parasitic infections.

**Discussion**

The fourfold pathogen reference ontology suite provides rigorous ontological representations more specific than IDO but more general than pathogen-specific domain-level ontologies. We maintain that our suite will facilitate consistency and coherence of domain-level ontologies that extend from it by following the hub-and-spoke design strategy and adherence to OBO Foundry principles, including the reuse of terms where possible from well-established domain ontologies, such as the Gene Ontology (GO) and the Protein Ontology (PRO). Of course, leveraging ontologies in the suite to realize improvements in data quality requires the subsequent implementation of these ontologies within research workflows, including their use in data curation, integration, and analysis. We consider our work here as laying further foundations for such implementations and subsequent empirical investigation, which is why domain experts have been actively involved in the ontology development process, contributing to term selection, definition refinement, and quality control. In addition to promoting accuracy, involving domain experts increases the likelihood of adoption in ontology-driven applications, where the extent to which ontologies in the suite can be used to improve data quality can be explored empirically.

As mentioned in the introduction, we take this work to be in the spirit of the OBO Foundry COB effort to construct a mid-level ontology that is "anchored in" BFO but which omits certain BFO commitments that have not traditionally been needed by biomedical ontology users, such as **two-dimensional spatial region** or **one-dimensional continuant fiat boundary**. COB classes have or will have corresponding BFO equivalencies, however, providing a mapping between the two when needed. As a consequence, though we have opted to leverage BFO for the development of each reference ontology in the suite, these mappings should ultimately allow for the use of either.

Our motivation for leveraging BFO in this work stems largely from BFO providing a more expressive semantics within which to disambiguate domain-level terms and relations. Our team found this invaluable when initiating ontology development. That should not, however, suggest that every ontology-driven application pulling from our suite must use all of BFO, IDO, and so on. Decisions, for example, on whether to leverage ontologies in the suite using BFO or COB will be answered based on specific application needs and implementations in software applications.

There are challenges emerging from our work worth highlighting. One significant issue is the need for continuous updates to these ontologies as new pathogens emerge and research priorities shift. The modularity of these ontologies can, in this context, be a blessing and a curse. On the one hand, modularity allows for updates to domain-level ontologies without necessarily requiring updates to reference-level ontologies they import. On the other hand, updates to reference ontologies may have significant impacts on ontologies that import them. Throughout, there remains a need for an organized and standardized mechanism to manage changes across multiple ontology levels. This includes ensuring that updates to terms, definitions, and relationships are consistently implemented across extensions and that new terms do not conflict with existing ones. Additionally, as new pathogen-specific ontologies continue to be developed, maintaining interoperability between them and existing ontologies under IDO requires ongoing coordination and shared governance.

Steps forward in this direction include having IDO, VIDO, BIDO, MIDO, and PIDO housed under a common development environment, such as the Infectious Disease Extension (https://github.com/infectious-disease-ontology-extensions) GitHub Organization, adopting the same ontology development standards. To that end, an active coordination effort is underway among the respective developer teams to bring together not only the ontologies comprising the suite, but all up-to-date versions of every extension as well, such as CIDO, IDOMAL, and so on. Stakeholders are encouraged to submit generalized terms to the IDO Core repository via GitHub

Issues, where a designated curation team evaluates whether the proposed term falls within IDO's scope during biweekly meetings. Similarly, in cases where a term submitted to, say, PIDO is broadly relevant across pathogen types, the PIDO team flags the term for discussion in biweekly IDO coordination meetings, where inclusion will be determined once consensus (or majority vote) has been reached between ontology engineers and subject-matter experts.

    Key to this coordination effort is the incorporation of continuous integration and deployment (CI/CD) best practices [91]. Presently, each ontology repository maintains standard production (for stable releases) and development (for ongoing work) branches and deploys the ROBOT tool [92] into respective workflows using GitHub actions, which allow for running integration tests to ensure data consistency and integrity, deploying a release to stakeholders, and deployment of artifacts to a staging environment for review. The ROBOT tool provides readymade unit tests and ontology quality control checks. Through our pipeline, contributors will fork the main repository they would like to edit, make updates, then submit a pull request to the development branch of that repository, which triggers actions to run quality control tests on their pull request. Following significant updates, the development branch will be merged into the production branch, generating a new release of the ontology, documentation, changelog, and so on. Prior to any release, stakeholders will be encouraged to review the staging environment to provide feedback. Following a release, the relevant ontology team will gather feedback from stakeholders, which will be prioritized for discussion during coordination meetings. Additionally, each team is drafting user-facing guides that include instructions for submitting term requests, criteria for determining the appropriate ontology to which a term should be submitted, and templates for providing minimum metadata (definition, examples, logical relations, source). These guides supplement issue and feature request templates, contributor guides, and github documentation already found in each repository.

The preceding provides lines along which we may address another challenge, namely the status of legacy extensions of IDO that were once active but are no longer maintained due to discontinued projects or shifting institutional support, such as IDOSCHISTO and IDOHIV [3]. These ontologies often contain valuable domain-specific content but may use outdated modeling practices, inconsistent identifiers, or deprecated upper ontology terms. For our part, we aim to provide robust mappings of their terms to reference-level classes within IDO, VIDO, BIDO, MIDO, or PIDO, following the strategy developed in [93], to preserve term provenance and definitions while maintaining alignment with IDO releases. These efforts will also be coordinated through our Infectious Disease Ontology Extensions GitHub Organization.

## Conclusion

A central conceit of this work is that rigorous, domain-expert driven, ontological representations more specific than IDO but more general than domain-level ontologies, provide guardrails within which domain-level ontologies can be constructed, saving researchers time and effort. In that respect, we view our work as analogous to the way Python libraries save software developers time and effort, as they are building applications for specific needs. We look forward to continued work with the broader ontology community, refining, expanding, and implementing these ontologies using bioinformatics tools.

## Declarations

**Ethics Approval and Consent to Participate:** not applicable.

**Consent for Publication:** not applicable.

**Availability of Data and Materials.** The artifacts reported in this work can be found on GitHub in the following locations:
- VIDO: https://github.com/infectious-disease-ontology-extensions/VIDO
- BIDO: https://github.com/infectious-disease-ontology-extensions/BIDO
- MIDO: https://github.com/infectious-disease-ontology-extensions/MIDO
- PIDO: https://github.com/infectious-disease-ontology-extensions/PIDO

**Competing Interests:** The authors declare that they have no competing interests

**Funding:** This paper was partially supported by the National Institute of Allergy and Infectious Diseases of the National Institutes of Health (award number U24AI171008).